\documentclass[sigconf]{acmart}
\usepackage{algorithm}
\usepackage{algorithmic}
\usepackage{amsmath}
\usepackage{amsthm}
\usepackage{float} 
\usepackage{centernot}
\usepackage{multirow} 
\usepackage{array}
\usepackage{subfigure}
\usepackage{xcolor}
\usepackage{color}
\usepackage{float}
\usepackage{xspace}

\AtBeginDocument{%
 }

\setcopyright{acmlicensed}
\copyrightyear{2018}
\acmYear{2018}
\acmDOI{XXXXXXX.XXXXXXX}

\acmConference[Conference acronym 'XX]{Make sure to enter the correct
 conference title from your rights confirmation emai}{June 03--05,
 2018}{Woodstock, NY}
\acmISBN{978-1-4503-XXXX-X/18/06}

\begin{document}

\title{Multi-Cause Deconfounding for Recommender Systems with Latent Confounders}


\author{Zhirong Huang}
\authornote{Both authors contributed equally to this research.}
\author{Shichao Zhang}
\authornote{Corresponding author.}
\affiliation{%
	\institution{Key Lab of Education Blockchain and Intelligent Technology, Ministry of Education}
	\institution{Guangxi Normal University}
	\city{Guilin}
	\state{Guangxi}
	\postcode{541004}
	\country{China}}
\affiliation{%
	\institution{Guangxi Key Lab of Multi-Source Information Mining and Security}
	\institution{Guangxi Normal University}
	\city{Guilin}
	\state{Guangxi}
	\country{China}}

\author{Debo Cheng}
\authornotemark[1]
\authornotemark[2]
\author{Jiuyong Li}
\author{Lin Liu}
\affiliation{%
	\institution{UniSA STEM, University of South Australia}
	\city{Mawson Lakes}
	\state{Adelaide}
	\country{Australia}}

\author{Guixian Zhang}
\affiliation{%
	\institution{China University of Mining and Technology}
	\city{Xuzhou}
	\state{Jiangsu}
	\postcode{221116}
	\country{China}}

\renewcommand{\shortauthors}{Zhirong Huang, Shichao Zhang, Debo Cheng et al.}
\newtheorem{owndefinition}{Definition}

\begin{abstract}
 
In recommender systems, various latent confounding factors (e.g., user social environment and item public attractiveness) can affect user behavior, item exposure, and feedback in distinct ways. These factors may directly or indirectly impact user feedback and are often shared across items or users, making them multi-cause latent confounders. However, existing methods typically fail to account for latent confounders between users and their feedback, as well as those between items and user feedback simultaneously. To address the problem of multi-cause latent confounders, we propose a \underline{m}ulti-\underline{c}ause \underline{d}e\underline{c}on\underline{f}ounding method for recommender systems with latent confounders (MCDCF). MCDCF leverages multi-cause causal effect estimation to learn substitutes for latent confounders associated with both users and items, using user behaviour data. Specifically, MCDCF treats the multiple items that users interact with and the multiple users that interact with items as treatment variables, enabling it to learn substitutes for the latent confounders that influence the estimation of causality between users and their feedback, as well as between items and user feedback. Additionally, we theoretically demonstrate the soundness of our MCDCF method. Extensive experiments on three real-world datasets demonstrate that our MCDCF method effectively recovers latent confounders related to users and items, reducing bias and thereby improving recommendation accuracy.
\end{abstract}

\ccsdesc[500]{Information systems~Collaborative filtering}

\keywords{Recommender system, Causal inference, Latent confounder}
\maketitle
\section{Introduction}
Recommender systems are designed to recommend the items that align with users' preferences from vast amounts of available items~\cite{singh2021recommender}. By providing accurate recommendations, these systems significantly improve user experience and generate higher commercial revenues for online service platforms. Therefore, recommender systems are widely used in various online services, including streaming media~\cite{gomez2015netflix}, e-commerce~\cite{shoja2019customer}, and other fields~\cite{ko2022survey}.

\begin{figure}[t]
\begin{center}
  \subfigure[]{
    \includegraphics[width=0.23\columnwidth]{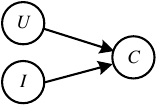}
    }
    \subfigure[]{
    \includegraphics[width=0.23\columnwidth]{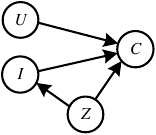}
    }
    \subfigure[]{
    \includegraphics[width=0.24\columnwidth]{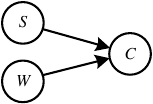}
    }
  \caption{Three causal graphs illustrate the assumptions underlying existing research. The variables are defined as follows: $U$ (user), $I$ (displayed items), $C$ (user feedback), $Z$ (item popularity), $S$ (user interest), and $W$ (user conformity). (a) The causal graph represents the traditional recommendation model; (b) the causal graph represents the PDA model; (c) the causal graph represents the DICE model.}
  \Description{}
  \label{Fig1} 
\end{center}
\end{figure}

Traditional recommendation methods are often based on collaborative filtering, which analyses user behaviour data to identify similarities between users and items to predict user preferences~\cite{wu2022survey,papadakis2022collaborative, zhao2022investigating}. For instance, Matrix Factorisation (MF)~\cite{koren2009matrix} algorithms predict user preferences and generate personalised recommendations by decomposing the user behaviour matrix into a product of low-dimensional user and item feature matrices. These methods typically rely on a key assumption: user behaviour data is unbiased, meaning that user feedback is generated solely when a user interacts with an item, without being influenced by other confounding factors. This assumption is often represented in causal directed acyclic graphs (DAGs)~\cite{pearl2009causality}, as shown in Figure~\ref{Fig1} (a).

In real-world scenarios, user-item interactions are frequently influenced by various confounding factors that beyond user preferences and item attributes~\cite{luo2024survey,chen2023bias}. For instance, interactions within recommender systems may be shaped by external influences such as item popularity or social environment, causing users' behaviour to deviate from their genuine preferences. Consequently, traditional recommendation methods risk introducing bias when predicting these preferences~\cite{gao2024causal}. Thus, accurately identifying and mitigating the effects of these confounding factors has become a significant challenge in contemporary recommender systems, where the goal is to capture users' true preferences more precisely.

To mitigate the effects of confounders in recommender systems, causal inference methods have been employed to mitigate the bias of specific confounders based on a defined causal DAG~\cite{wang2020causal}. For example, Zhang et al.\cite{zhang2021causal} constructed a causal graph (see Figure~\ref{Fig1} (b)) to identify item popularity by analysing its impact on user behaviour from the item’s perspective. They developed the Popularity-bias Deconfounding and Adjusting (PDA) method to mitigate bias by applying \textit{do-calculus}~\cite{pearl2009causality} to adjust for item popularity as a confounder. Similarly, Zheng et al.\cite{zheng2021disentangling} introduced another causal graph to model the influence of conformity on user behaviour, treating it as a combination of conformity-driven and interest-driven actions (see Figure~\ref{Fig1} (c)). They proposed the Disentangling Interest and Conformity with Causal Embedding (DICE) method, which reduces conformity bias by constructing specific training sets. These causal graph-based debiasing methods rely on the assumption that specific confounders can be identified within the user behaviour data~\cite{cai2024mitigating}. While these methods do successfully identify specific confounders (e.g., item popularity), many confounders remain unobserved in real-world scenarios (i.e., latent confounders) within user behavior data, such as a user’s social environment or an item’s public attractiveness. This limits the applicability and effectiveness of these debiasing methods.

\begin{figure}[t]
\begin{center}
  \subfigure[]{
    \includegraphics[width=0.65\columnwidth]{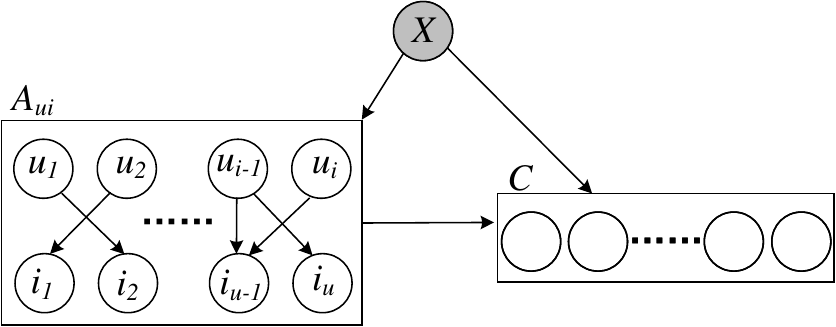}
    }
    \subfigure[]{
    \includegraphics[width=0.65\columnwidth]{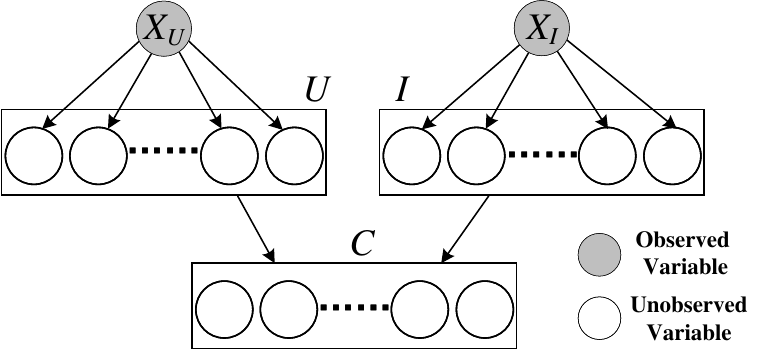}
    }
  \caption{Two causal graphs illustrate multi-cause debiasing methods. The variables are defined as follows: $U$ (users), $I$ (displayed items), $C$ (user feedback), $A_{ui}$ (user exposure, i.e., the user's implicit interaction data with the item), $X$ (the set of latent confounders), $X_I$ (latent confounders for items), and $X_U$ (latent confounders for users). (a) The first causal graph represents the DeConfounder method; (b) the second causal graph represents our proposed MCDCF method.}
  \Description{}
\label{Fig2} 
\end{center}
\end{figure}

To mitigate the effects of latent confounders in recommender systems, multi-cause debiasing methods have been developed based on the idea of learning substitutes for latent confounders in multi-cause settings~\cite{wang2019blessings}. For example, Wang et al.~\cite{wang2020causal} framed the recommendation problem as a multi-cause problem, arguing that latent confounders simultaneously affect multiple treatment variables (i.e., user exposure) and outcome variables (i.e., user feedback), as shown in Figure~\ref{Fig2} (a). They proposed the DeConfounder method, which learns substitute confounders from user exposure data to mitigate the influence of latent confounders. Building on the success of DeConfounder, several variant approaches have emerged, such as the Identifiable Deconfounder (iDCF)~\cite{zhang2023debiasing} and the Deep Deconfounded Recommender (Deep-Deconf)~\cite{zhu2024deep}. These variants enhance the characterisation of substitute confounders by incorporating additional information, such as the user's income. These methods provide a coarse analysis of latent confounders in user-item interactions, overlooking the distinctions between confounders that impact the user side versus those that affect the item side. For instance, a user’s social environment may influence the user but not the item, leading to biased substitute confounders. Additionally, these methods often rely on hard-to-access information, such as the user's income, due to privacy concerns, limiting their practical applicability. Thus, effectively mitigating the effects of latent confounders in recommender systems remains a significant challenge. 

To address this challenge, we model the recommendation task as a multi-cause problem and propose a causal graph to analyse the latent confounders affecting the user side and item side, respectively, as shown in Figure~\ref{Fig2} (b). Moreover, we analyse the impact of latent confounders on user behaviour from the user and item sides separately instead of roughly analysing the effect of confounders on user-item interaction only. We propose a novel \underline{m}ulti-\underline{c}ause \underline{d}e\underline{c}on\underline{f}ounding for recommender systems, referred to as \textbf{MCDCF}. MCDCF treats the multiple items associated with a user and the multiple users associated with an item as treatment variables, enabling it to learn substitutes for the two latent confounders that affect the causal relationship between users and user feedback, as well as the relationship between items and user feedback.  These learned substitute confounders are then utilized, along with user and item features, to perform click prediction (i.e., predicting user feedback) and recommend candidate items to the user.  In summary, our contributions are as follows:
\begin{itemize}
  \item We propose a multi-cause framework for recommender systems that analyses the impact of latent confounders on user behaviour from both the user and item sides.
  \item We propose an end-to-end multi-cause deconfounding for recommender systems with latent confounders, which directly learns substitute confounders from user behaviour data on both the item and user sides, without requiring additional information. 
  \item We conducted extensive experiments on three real-world datasets to demonstrate the effectiveness and superiority of the substitute confounders learned by MCDCF method.
\end{itemize}

\section{Related Work}
This section introduces debiasing methods based on causal inference in recommender systems. Depending on the type of confounders, we discuss two classes of debiasing methods: those for observed confounders and those for latent confounders.

\subsection{Debiasing Methods for Observed Confounders} 
In recent years, numerous debiasing solutions based on causal inference have emerged~\cite{zhu2024mitigating, wei2021model}. Researchers have discovered that confounders can bias recommender systems, resulting in inaccurate predictions of user preferences. The most straightforward approach to mitigating bias is to eliminate observed confounders, which are typically identifiable from user behavioural data, such as item popularity. Early research applied the Inverse Propensity Score (IPS)~\cite{bottou2013counterfactual} to address popularity bias by using the inverse of item popularity as the IPS to balance the importance of popular and less popular items. However, IPS methods are highly dependent on item popularity and suffer from high variance, which has prompted the development of various variant methods~\cite{wang2021non,schnabel2016recommendations} to mitigate this issue. 

With advances in causal inference, researchers have proposed debiasing methods based on causal graphs~\cite{zheng2021disentangling, zhang2021causal,zhao2023disentangled,he2023addressing}. These approaches utilize domain knowledge to analyse the data generation mechanisms underlying user behaviour and employ causal graphs to identify observed confounders. These confounders are addressed by developing models to handle specific biases, such as popularity and conformity bias. However, many real-world scenarios involve confounders that are not directly observed (e.g., user's location) and may be hidden within behavioural patterns. As a result, the presence of latent confounders limits the applicability of IPS and causal graph-based debiasing methods.

\subsection{Debiasing Methods for Latent Confounders}

To address latent confounders in recommender systems, researchers have explored debiasing methods based on instrumental variables (IVs)~\cite{pearl2009causality} and multiple causal inference~\cite{wang2019blessings}. These methods aim to account for latent confounders by incorporating additional information. For example, Si et al.~\cite{si2022model} used self-collected user search data as an IV to mitigate the effects of latent confounders by decomposing user behaviour into causal and non-causal components. However, the existing IV approach~\cite{si2023enhancing, si2023search, si2022model} requires a predefined IV~\cite{cheng2023causal}, which makes it challenging to identify a convincing IV from observed recommender system data.

To address these limitations, researchers have explored multiple causal inference-based solutions~\cite{wang2020causal,zhu2024deep,zhang2023debiasing}. These methods combine observed recommendation data with additional user information to learn substitute latent confounders, conditioning on these substitutes to mitigate the effects of confounders. While these approaches have demonstrated some success, they exhibit notable drawbacks. First, they provide only a rough analysis of the impact of latent confounders on user feedback, failing to account for differences between user-side and item-side confounders. Second, these methods often rely on additional user information, which is challenging to obtain in practice due to privacy-preserving policies, thereby further limiting their applicability. In our MCDCF method, we aim to address these shortcomings in the context of recommender systems.

\section{Problem Formulation}
Recommender systems aim to predict user preferences by analysing user behaviour data to estimate a user's feedback on items. The user behaviour data typically consists of a user set $U$ and an item set $I$. Sample pairs from the user and item sets form the user behaviour data: $D = \{(u, i) \mid u \in U, i \in I\}$, where each pair $(u, i)$ represents user feedback $C$, indicating that user $u$ has interacted with item $i$. Traditional recommender systems often assume that user behaviour data $D$ is unbiased, meaning users interact only with items that align with their preferences. In this context, the prediction model $Y$ can be expressed as:
\begin{equation}
  Y = p(C\mid U,I).
\end{equation}

Thus, user feedback $C$ is predicted based on the users $U$ and the items they are exposed to $I$. However, in real-world scenarios, user feedback is influenced by more than just users and items; there are many latent confounders. These latent confounders are typically unobserved, making it highly challenging to recover them from the data. Recent approaches have proposed modelling the recommendation problem as a multiple causal inference problem, using the user's exposure state $A_{ui}$ as the treatment variable and learning the posterior distribution of $A_{ui}$ as a substitute for the latent confounders $X$. The prediction model $Y$ can then be expressed as:
\begin{equation}
  Y = \mathbb{E}[p(C\mid U, I, X)].
\end{equation}

However, the substitute for latent confounders $X$ recovered by existing methods is biased because they only roughly account for the effect of confounders on the user's exposure state $A_{ui}$, without differentiating between user-side and item-side latent confounders. For instance, consider a social environment as a latent confounder on the user side, where a user is influenced by peers to watch an action movie despite having no prior interest in the genre. In this case, the latent confounders affect the user’s preferences and subsequent feedback, but they have no impact on the item itself. In this work, we aim to eliminate this bias in recommender systems by separately recovering latent confounders from both the user and item sides. Our problem definition is outlined below.

\begin{figure}[h]
    \centering
    \includegraphics[width=0.8\linewidth]{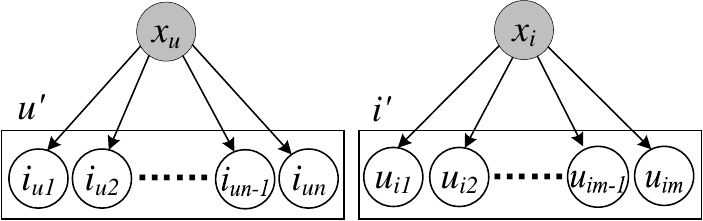}
    \caption{ A partial causal graph illustrating the latent confounders between users and items. $u^{\prime}$: substitute representation of user $u$, $i^{\prime}$: substitute representation of item $i$, $i_{im}$: items user $u$ has interacted with, $u_{um}$: users who have interacted with item $i$.}
    \label{local_model}
    \Description{}
\end{figure}

\begin{owndefinition}[problem definition] 
We assume that latent confounders simultaneously affect multiple causes. For each user $u$ and item $i$, the corresponding substitute latent confounders are denoted as $x_u$ and $x_i$, where $u \in U$, $i \in I$, $x_u \in X_U$, and $x_i \in X_I$. We recover $x_u$ and $x_i$ from the substitute representations $u^\prime$ and $i^\prime$, where $u^\prime$ and $i^\prime$ are formed by the interactions of user $u$ and item $i$, respectively. Specifically, $u^\prime = [i_{u1}, i_{u2}, ..., i_{un}] \in \mathbb{R}^{n \times d}$ and $i^\prime = [u_{i1}, u_{i2}, ..., u_{im}] \in \mathbb{R}^{m \times d}$, where $n$ and $m$ represent the numbers of selected interactions, as illustrated in Figure~\ref{local_model}. Our goal is to recover the substitutes $X_U$ for users and $X_I$ for items from the behaviour data $D$, and then combine them to perform click prediction. Thus, our final model is: $Y = \mathbb{E}[p(C \mid U, I, X_U, X_I)]$.
\end{owndefinition}

\begin{figure*}[htbp]
  \centering
  \includegraphics[width=0.6\linewidth]{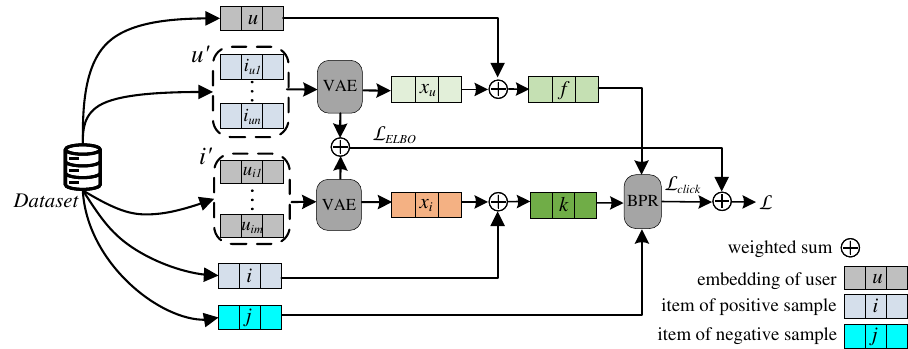}
  \caption{The overall architecture of MCDCF, which involves two core steps. First, the MCDCF learns substitute latent confounders $x_u$ and $x_i$ representing users and items from the user behaviour dataset. Second, $x_u$ and $x_i$ are combined with their corresponding users $u$ and items $i$, resulting in fused features $f$ and $k$ that target the positive sample data, followed by model optimisation. This process enables MCDCF to generate debiased recommendations.}
  \Description{}
  \label{framework}
\end{figure*}

\section{The Proposed MCDCF Method}
Our proposed multi-cause deconfounding method (MCDCF) is developed based on the causal graph shown in Figure~\ref{Fig2} (b). The method consists of two main steps. First, we recover the set of substitute latent confounders, $X_U$ for users and $X_I$ for items, from the user behaviour data. The second step involves conditioning on these substitute confounders to predict user feedback.

\subsection{The Proposed Causal Graph for MCDCF}
In this work, we propose a causal graph, as shown in Figure~\ref{Fig2} (b), to model the recommendation problem as a multi-cause problem. Unlike DeConfounder and its variants, our approach accounts for the influence of latent confounders (i.e., $X_U$ and $X_I$) on both users $U$ and items $I$. Specifically, we treat the multiple items associated with user $u$ and the multiple users associated with item $i$ as treatment variables, as illustrated in Figure~\ref{local_model}. We then learn the substitute confounders $x_u$ and $x_i$ from $u^\prime$ and $i^\prime$, which affect user $u$ and item $i$, respectively.
 
 (1) We assume that the latent confounder $x_u$ on the user side is the social environment, which affects all users but to varying degrees. For some users, the social environment may alter their preferences, thereby influencing their feedback, as represented by the $X_U \rightarrow U \rightarrow C$ path in Figure~\ref{Fig2}(b). Simultaneously, social environment may lead a user to watch movies she/he has never seen before without changing her/his actual interest, corresponding to the $X_U \rightarrow C$ path in Figure~\ref{Fig2} (b). However, the social environment does not affect item attributes. 
 
 (2) On the item side, we assume the latent confounder $x_i$ is the item's public attractiveness, such as endorsements from public figures for certain movies. This endorsement increases the movie's exposure, making it more likely to be chosen by users, as shown by the $X_I \rightarrow I \rightarrow C$ path. Additionally, the endorsement may cause some users to select the movie regardless of its exposure or quality, represented by the $X_I \rightarrow C$ path. Similarly, item public attractiveness does not influence user attributes.  

 These examples demonstrate that user and item sides possess distinct confounders. Therefore, it is essential to separately analyse the effects of user-side and item-side confounders on user behaviour.

\subsection{Recovering the Substitutes of Latent Confounders}
The key to achieving $E[p(C\mid U, I, X_U, X_I)] = E[p(C)]$ lies in recovering the sets of latent confounders $X_U$ and $X_I$ for users and items, respectively. Specifically, for each user $u$ and item $i$, we aim to learn their posterior distributions $p(x_u\mid u^{\prime})$ and $p(x_i\mid i^{\prime})$ as substitutes for latent confounders, as shown in Figure~\ref{local_model}. Here, $u^{\prime}$ and $i^{\prime}$ represent the substitute embeddings of user $u$ and item $i$, respectively, composed of their interactions with items and users (where latent confounders are implicitly present in the observed data). Specifically, $u^\prime = [i_{u1}, i_{u2}, ..., i_{un}]$ and $i^\prime = [u_{i1}, u_{i2}, ..., u_{im}]$, where $n$ and $m$ are the numbers of selected interactions.  It is important to note that $u^{\prime}$ and $i^{\prime}$ encompass multiple causes. As they are derived from selected interactions, they implicitly contain user feedback information (i.e., C in Figure~\ref{Fig2} (b)). Our aim is to recover the substitute confounders $x_u$ and $x_i$ from these multiple causes. 

Recently, Variational Autoencoders (VAE)~\cite{kingma2013auto} have achieved significant success in causal representation learning. In this work, we leverage VAE to learn substitute latent confounders from user behaviour data. We adopt the standard VAE by assuming that the prior distributions $p(x_u)$ and $p(x_i)$ of each user and item features obey a Gaussian distribution:
\begin{equation}
  p(x_u) \sim \mathcal{N}(x_u \mid 0, \mathbf{I}),
\end{equation}
\begin{equation}
   p(x_i) \sim \mathcal{N}(x_i \mid 0, \mathbf{I}),
\end{equation}
where $\mathbf{I}$ denotes the identity matrix. Meanwhile, we recover $x_u \in \mathbb{R}^d$ and $x_i\in \mathbb{R}^d$ using two independent VAE models, as shown in Figure~\ref{framework}. During the inference phase of the VAE, the encoders $q(x_u\mid u^{\prime})$ and $q(x_i\mid i^{\prime})$ are utilised as variational posteriors for $x_u$ and $x_i$:
\begin{equation}
  q(x_u\mid u^{\prime})=\mathcal{N}\left(x_u\mid\mu = mean(\widehat{\mu}_{x_{u}}),\sigma^2 = mean(\widehat{\sigma}_{x_{u}}^2) \right),
\end{equation}
\begin{equation}
  q(x_i\mid i^{\prime})=\mathcal{N}\left(x_i\mid\mu = mean(\widehat{\mu}_{x_{i}}),\sigma^2 = mean(\widehat{\sigma}_{x_{i}}^2) \right),
\end{equation}
where $\widehat{\mu}_{x_{u}}$, $\widehat{\mu}_{x_{i}}$ and $\widehat{\sigma}_{x_{u}}^2$, $\widehat{\sigma}_{x_{i}}^2$ are the means and variances parameterised by neural networks, and $mean()$ represents an averaging operation. $x_u$ and $x_i$ are multi-cause confounders that affect multiple causes simultaneously (i.e., $u^\prime$ and $i^\prime$). Therefore, we extract the latent features shared in $u^\prime$ and $i^\prime$ through averaging operations on the mean and variance in the encoder. We use Kullback-Leibler~\cite{van2014renyi} divergence to ensure that $q(x_u\mid u^\prime)$ and $q(x_i\mid i^\prime)$ approximate the prior distributions $p(x_u)$ and $p(x_i)$:
\begin{equation}
\begin{split}
D_{KL}(q(x_u \mid u^{\prime}) \ \mid p(x_u)) 
 =\int q(x_u \mid u^{\prime}) \log \frac{q(x_u \mid u^{\prime})}{p(x_u)} \, dx_u,
\end{split}
\end{equation}

\begin{equation}
\begin{split}
D_{KL}(q(x_i \mid i^{\prime}) \ \mid p(x_i))  
=\int q(x_i \mid i^{\prime}) \log \frac{q(x_i \mid i^{\prime})}{p(x_i)} \, dx_i.
\end{split}
\end{equation}

During the generation phase of the VAE, we use Gaussian sampling to obtain samples $x_u$ and $x_i$:
\begin{equation}
    x_u \sim q(x_u\mid u^{\prime}),
\end{equation}
\begin{equation}
    x_i \sim q(x_i\mid i^{\prime}).
\end{equation}

The process of generating data using substitute latent confounders $x_u$ and $x_i$ can be represented as follows:
\begin{equation}
    \bar{u} =  Decoder(x_{u}),
\end{equation}
\begin{equation}
    \bar{i} = Decoder(x_{i}).
\end{equation}

In this way, we can approximate the process of $x_u \rightarrow u^\prime$ and $x_i \rightarrow i^\prime$ in the causal graph of Figure~\ref{local_model}, and further ensure the $x_u$ and $x_i$ recovered from user's behavioural data of reliability:
\begin{equation}
  \mathcal{L}_u=\mathbb{E}_{q(x_u \mid u^\prime)} \left[ \| u - \bar{u} \|^2 \right],
\end{equation}
\begin{equation}
  \mathcal{L}_i=\mathbb{E}_{q(x_i \mid i^\prime)} \left[ \| i - \bar{i} \|^2 \right]. 
\end{equation}

Finally, we use the evidence lower bound (ELBO) to optimise the parameters:
\begin{equation}
  \mathcal{L}_{ELBO}^U = \mathcal{L}_u - D_{KL}(q(x_u \mid u^\prime),
\end{equation}
\begin{equation}
  \mathcal{L}_{ELBO}^I = \mathcal{L}_i - D_{KL}(q(x_i \mid i^\prime),
\end{equation}
\begin{equation}
  \mathcal{L}_{\text{ELBO}} = \frac{1}{2} \left( \mathcal{L}_{\text{ELBO}}^U + \mathcal{L}_{\text{ELBO}}^I \right).
\end{equation}

By maximising the $\mathcal{L}_{\text{ELBO}}$, we can obtain posteriors $p(x_u\mid u^\prime)$ and $p(x_i\mid i^\prime)$ as the substitute latent confounders. In summary, we treat the items interacted with by user $u$ as multiple causes to recover the user's substitute confounder, $x_u$. Similarly, the users interacted with by item $i$ are considered multiple causes to recover its substitute confounder, $x_i$. We then use $x_u$ and $x_i$ to reconstruct $u$ and $i$, ensuring the validity of both factors.

\subsection{Debiased Implicit Feedback Recommendation}
To achieve our goal $Y = \mathbb{E}[p(C\mid U, I, X_U, X_I)]$, We need to use marginalisation to calculate the expected probability of user feedback $C$:
\begin{equation}
\begin{split}
  \mathbb{E}[p(C\mid U, I, X_U, X_I)] = &\int \int P(C \mid U, I, X_U, X_I) \\&\cdot P(X_U \mid U^\prime) \cdot P(X_I \mid I^\prime) \, dX_U \, dX_I,
\end{split}
\end{equation}
where the $U^\prime$ and $I^\prime$ represent the sets of $u^\prime$ and $i^\prime$, which are approximated by $U$ and $I$, respectively.

Although theoretically we need to integrate all possible values of $X_U$ and $X_I$, in practical implementations, it is often difficult to solve this integral explicitly. Therefore, we adopt an approximate method based on Monte Carlo sampling to estimate this integral. By sampling from the posterior distributions $p(X_U\mid U^\prime)$ and $p(X_I\mid I^\prime)$ of $X_U$ and $X_I$, we can generate multiple samples of $X_U$ and $X_I$ and compute the value of $P(C\mid U,I,X_U,X_I)$ for each sample:
\begin{equation}
  \mathbb{E} [ P ( C \mid U , I , X_U , X_I ) ] \approx \frac { 1 } { N } \sum _ { n = 1 } ^ { N } P ( C \mid U , I , X_{U_n} , X_{I_n} ),
\end{equation}
where $X_{U_n}$ and $X_{I_n}$ are the $n$-th samples of $X_U$ and $X_I$ sampled from $P(X_U\mid U^\prime)$ and $P(X_I\mid I^\prime)$ and $N$ is the number of samples. In practice, to simplify the computation, we only perform a single sampling, and $P(X_U\mid U^\prime)$ and $P(X_I\mid I^\prime)$ are computed by the VAE.

We model user features and item features fused with corresponding substitute confounders using a simple additive model, to more fully capture the explicit and implicit factors that influence user behaviour:
\begin{equation}
  f = u + \alpha * x_{u},
\end{equation}
\begin{equation}
  k = i + \beta * x_{i},
\end{equation}
where $\alpha$ and $\beta$ are the hyperparameters controlling the weights. This design enables the model to consider both the explicit features of users and items, as well as the latent confounders behind them, thereby enhancing the ability to match users with items. Note that the confounders affect the user behaviour, i.e., the positive samples of the embodied user behaviour data, i.e., the sample pairs $(u, i)$ consisting of the user and the selected item, and thus the fused features $f$ and $k$ target the positive sample data, as shown in Figure~\ref{framework}.

Click prediction is a core task in recommender systems that usually relies on item ordering. To enhance the recommender system's ability to predict user clicking behavior, we employ the Bayesian Personalized Ranking (BPR)~\cite{rendle2012bpr} loss to optimize this ordering:
\begin{equation}
\begin{split}
\mathcal{L}_{click} = -\sum_{\left(u,i,j\right) \in D} \ln \sigma \big( 
\left<f_u, k_{i} \right> - \left<f_u, i_j\right> \big),
\end{split}
\end{equation}
where $<\cdot,\cdot>$ denotes the inner product operation, $\sigma$ is the sigmoid activation function, and $j$ is a negative sample, an item in the dataset that the user $u$ did not interact with, and we use the Popularity based Negative Sampling with Margin (PNSM)~\cite{zheng2021disentangling} method to select the negative samples. It is worth noting that when we select the negative samples, we choose to draw from the raw features of the items $i$ instead of $k$. The reason for this design is that the negative samples represent the items that the user has not interacted with, and the latent confounder $x_i$ does not play a decisive role in these items. In contrast, raw feature $i$ is more suitable for describing these uninteracted items. Finally, we combine the click loss and the ELBO to obtain the final loss function:
\begin{equation}
  \mathcal{L}=\mathcal{L}_{click} + \mathcal{L}_{ELBO}. 
\end{equation}

In the inference phase of MCDCF, we use $F$ and $K$ to calculate the probability of user feedback $C$, which can be modelled by the following equation:
\begin{equation}
   P ( C \mid U , I , X_{U} , X_{I} ) = \ln\sigma<F, K>.
\end{equation}
 
In summary, we first use VAE to learn the posterior distribution of latent confounders in users and items, which is close to the prior distribution (i.e., the distribution of true latent confounders). We obtain substitute latent confounders $X_U$ and $X_I$ through Monte Carlo sampling in the learned posterior distribution. 
Then, we condition on $U$ and $I$ and the corresponding $X_U$ and $X_I$ to achieve debiased recommendation, that is, simulate $\{U, I, X_U, X_I\} \rightarrow C$ in the Figure~\ref{Fig2} (b).

\subsection{Theoretical Analysis of MCDCF Method}
Traditional recommender systems (e.g., MF) usually learn models by training on large amounts of user behavioural data to estimate $p(C\mid U, I) \sim p(C)$, where $C$ represents user feedback, and $U$ and $I$ represent user and item representations, respectively. This approach attempts to model the probability of user feedback $C$ based on the given $U$ and $I$. However, such an estimation of $p(C\mid U, I)$ can be biased due to latent confounders that affect $U$, $I$, and $C$, resulting in $p(C\mid U, I) \neq p(C)$, as shown in Figure~\ref{Fig2} (b). The factors affecting $C$ are $U$, $I$, $X_U$, and $X_I$. By recovering the latent confounders $X_U$ and $X_I$ in terms of users and items, and conditioning $U$, $I$, $X_U$ and $X_I$ conditionally, then $\mathbb{E}[p(C\mid U, I, X_U, X_I)] = \mathbb{E}[p(C)]$\footnote{We obtain substitute confounders $X_U$ and $X_I$ by approximating the posterior distributions of $U$ and $I$. Since $X_U$ and $X_I$ are distributions rather than specific values, the conditional probabilities cannot be computed as in the traditional model. Still, rather, $X_U$ and $X_I$ should be marginalised to obtain the expected probability $\mathbb{E}[p(C)]$ of the user's feedback $C$.}. We will show why MCDCF can achieve $\mathbb{E}[p(C\mid U, I, X_U, X_I)] = \mathbb{E}[p(C)]$ in two ways: firstly, whether or not $X_U$ and $X_I$ learnt by MCDCF are valid; and secondly, to justify $\mathbb{E}[p(C\mid U, I, X_U, X_I)] = \mathbb{E}[p(C)]$.

First, prior research~\cite{cheng2023causal, scholkopf2022causality} has demonstrated that VAE is capable of learning latent variable distributions from observed data. Accordingly, we assume that the substitute confounders $x_u$ and $x_i$ in MCDCF capture the distributions of multiple causes (i.e., $u^\prime$ and $i^\prime$) of the corresponding user $u$ and item $i$. The implication, then, is that the multiple causes in $u^\prime$ and $i^\prime$ are conditionally independent when conditioned on $x_u$ and $x_i$. If there are still unobserved confounders in the user behaviour data affecting $u^\prime$ and $i^\prime$ at this point, then according to the \textit{d-separation}~\cite{pearl2009causality} theorem in causal inference (we provide the causal inference related knowledge involved in our work in Appendix~\ref{Preliminaries}), the multiple causes in $u^\prime$ and $i^\prime$ are interdependent.  This contradicts our assumptions. By proof by contradiction, we can show that $x_u$ and $x_i$ capture substitute confounders, and therefore they are valid.
At the same time, we can deduce that $X_U$ and $X_I$ are valid because they are sets of substitute confounders $x_u$ and $x_i$.

Second, referring to the causal graph depicted in Figure~\ref{Fig2} (b), conditioning on $X_U$ blocks the backdoor path $X_U \rightarrow C$, thereby removing the confounding influence of $X_U$ on the causal effect of $U$ on $C$. Similarly, conditioning on $X_I$ blocks the backdoor path $X_I \rightarrow C$, isolating the causal effect of $I$ on $C$. As a result, when conditioned on $U, I, X_U$, and $X_I$, the equality $\mathbb{E}[p(C\mid U, I, X_U, X_I)] = \mathbb{E}[p(C)]$ holds, validating the effectiveness of the MCDCF method in eliminating bias introduced by latent confounders.

\section{Experiments}
In this section, we conduct a series of experiments on three real-world datasets to evaluate the recommendation and debiasing performance of the proposed MCDCF method.

\subsection{Experimental Settings}
We describe the dataset, baseline, and experimental parameters used in the experiment.

\noindent \paragraph{Datasets.} We conducted experiments on three real-world datasets: Douban-Movie~\cite{zhao2022popularity}, KGRec-Music~\cite{oramas2016sound}, and Amazon-Art~\cite{hou2024bridging}, which represent three common recommendation scenarios: video, music, and e-commerce, respectively. These datasets are derived from actual recommender systems and contain rich user rating information (i.e., user feedback). We binarised the datasets based on user ratings, converted them into datasets with implicit feedback, assigned 1 to 5-star ratings and 0 to the rest of the rating values, and applied 10-kernel filtering to the data. Table~\ref{Table 1} presents the statistical information of the preprocessed dataset. Detailed information on the preprocessing of the datasets can be found in Appendix~\ref{Preprocessing}.

\begin{table}[t]   
  \centering
    \caption{\centering Statistics of datasets.}
  \label{Table 1}
  \begin{tabular}{cccc} \hline
    Dataset & User \# & Item \# & Interaction \# \\ \hline
    Douban-Movie & 6,809 & 1,5012 & 173,766 \\
    KGRec-Music & 5,199 & 7,672 & 741,851 \\
    Amazon-Art & 15,139 & 29,836 & 265,594 \\
    \hline
  \end{tabular}

\end{table}

\noindent \paragraph{Baselines.} In our experiments, we will compare this with MF and its causal debiased expansion method:
\begin{itemize}
  \item \textbf{MF}~\cite{koren2009matrix}: The MF is a widely used collaborative filtering method in recommender systems that decomposes the user-item interaction matrix into two low-dimensional matrices representing the implicit features of users and items. Its goal is to reconstruct users' preferences for items from the product of these matrices.
  \item \textbf{IPS}~\cite{schnabel2016recommendations}: The Inverse Propensity Score (IPS) reduces the impact of popularity or exposure on recommendation results by assigning each user-item interaction a weight that is the inverse of the probability of that interaction occurring.
  \item \textbf{IPS-C}~\cite{bottou2013counterfactual}: This method is based on the IPS method and reduces the variance of the IPS values by maximising the truncation of the IPS values. IPS-C prevents excessive weighting by setting an upper limit, resulting in more stable estimates.
  \item \textbf{IPS-CN}~\cite{gruson2019offline}: This method further adds a normalisation operation to IPS-C, so that all IPS weights are normalised within a certain range.
  \item \textbf{IPS-CNSR}~\cite{gruson2019offline}: This method is based on IPS-CN with the addition of smoothing and renormalisation steps to obtain a smoother output.
  \item \textbf{DICE}~\cite{zheng2021disentangling}: This method is a recommendation model that separates users' interests from their follower behaviour. Through causal embedding and decoupled representation learning, DICE effectively distinguishes users' true interests from their follower behaviours for popular items.
  \item \textbf{DCCL}~\cite{zhao2023disentangled}: The method is based on the idea of causal decoupling of DICE, and enhances the learning of the representation of causal embedding through comparative learning, which enables the model to better distinguish between user interest and follower effect during the training process.
\end{itemize}

\paragraph{Metrics.} We assess the Top-K recommendation performance in the context of \textit{implicit feedback}, a prevalent scenario in recommender systems. The Top-K recommendations represent the K items the system identifies as most relevant or appealing to the user, where K specifies the number of items included in the recommendation list. We use three frequently applied evaluation metrics: \textit{Recall}, \textit{Hit Rate} (\textit{HR}), and \textit{NDCG}, to evaluate all methods. The reported experimental results represent the optimal performance achieved by each method under its respective parameter settings. 

Due to space constraints, we present the specific \textit{experimental parameter settings} in Appendix~\ref{Parameter}.

\begin{table*}[htbp]
\centering
\caption{The performance of all methods on the three real-world datasets. The best results are highlighted in bold, and the second-best results are underlined. }
\label{Comparison}
\setlength{\tabcolsep}{1.2mm}{ 
\begin{tabular}{ccr|ccccccccc}
\hline
Dataset&   \multicolumn{2}{c|}{Metric}& MF & IPS    & IPS-C  & IPS-CN & IPS-CNSR & DICE   & DCCL   & MCDCF  & Imp.  \\ \hline
\multirow{6}{*}{Douban-Movie} & \multirow{3}{*}{Topk=20} 
& Recall $\uparrow$ & 0.0201   & 0.0165 & 0.0155 & 0.0205 & \underline{0.0223}   & 0.0218 & 0.0197 & \pmb{0.0325} & 45.74\% \\
&& HR $\uparrow$     & 0.0664   & 0.0539 & 0.0536 & 0.0681 & 0.0745   & \underline{0.0755} & 0.0634 & \pmb{0.1110} & 47.02\% \\
& & NDCG $\uparrow$   & 0.0126   & 0.0097 & 0.0099 & 0.0135 & 0.0142   & \underline{0.0145} & 0.0120 &\pmb{ 0.0202} & 39.31\% \\ \cline{2-12} 

& \multirow{3}{*}{Topk=50} 
& Recall $\uparrow$ & 0.0351   & 0.0278 & 0.0268 & 0.0352 & \underline{0.0386}   & 0.0384 & 0.0323 & \pmb{0.0570} & 47.67\% \\
& & HR $\uparrow$     & 0.1162   & 0.0937 & 0.0909 & 0.1153 & 0.1244   & \underline{0.1281} & 0.1084 & \pmb{0.1820} & 42.08\% \\
& & NDCG $\uparrow$   & 0.0172   & 0.0132 & 0.0133 & 0.0179 & 0.0192   & \underline{0.0194} & 0.0159 & \pmb{0.0275} & 41.75\% \\ \hline

\multicolumn{1}{c}{\multirow{6}{*}{Amazon-Art}} & \multirow{3}{*}{Topk=20}
& Recall $\uparrow$ & 0.0090 & 0.0058 & 0.0054 & 0.0092 & 0.0099 & \underline{0.0141} & 0.0112 & \pmb{0.0247} & 75.18\% \\ \multicolumn{1}{c}{}         
& & HR $\uparrow$     & 0.0299   & 0.0190 & 0.0204 & 0.0308 & 0.0323   & \underline{0.0437} & 0.0388 & \pmb{0.0747} & 70.94\% \\
\multicolumn{1}{c}{}          
& & NDCG $\uparrow$   & 0.0048   & 0.0031 & 0.0032 & 0.0049 & 0.0052   & \underline{0.0080} & 0.0064 & \pmb{0.0144} & 80.00\% \\ \cline{2-12}

\multicolumn{1}{c}{}          & \multirow{3}{*}{Topk=50} 
& Recall $\uparrow$ & 0.0180   & 0.0116 & 0.0127 & 0.0183 & 0.0195   & \underline{0.0275} & 0.0217 & \pmb{0.0442} & 60.73\% \\ \multicolumn{1}{c}{}          
& & HR $\uparrow$     & 0.0579   & 0.0384 & 0.0436 & 0.0591 & 0.0628   & \underline{0.0844} & 0.0710 & \pmb{0.1286} & 52.37\% \\ \multicolumn{1}{c}{}          
& & NDCG $\uparrow$   & 0.0073   & 0.0048 & 0.0052 & 0.0075 & 0.0079   & \underline{0.0117} & 0.0093 & \pmb{0.0199} & 70.09\%  \\ \hline

\multirow{6}{*}{KGRec-Music}  & \multirow{3}{*}{Topk=20} 
& Recall $\uparrow$ & 0.0883   & 0.0728 & 0.0792 & \underline{0.1130} & 0.1011   & 0.0995 & 0.0899 & \pmb{0.1246} & 10.27\% \\
&  & HR $\uparrow$     & 0.8602   & 0.8017 & 0.8284 & \underline{0.9156} & 0.8909   & 0.8877 & 0.8642 & \pmb{0.9315} & 1.74\%  \\
& & NDCG $\uparrow$   & 0.1356   & 0.1116 & 0.1223 & \underline{0.1726} & 0.1546   & 0.1555 & 0.1397 & \pmb{0.1944} & 12.63\% \\ \cline{2-12} 
& \multirow{3}{*}{Topk=50} 
& Recall $\uparrow$ & 0.1597   & 0.1361 & 0.1482 & \underline{0.1988} & 0.1824   & 0.1799 & 0.1614 & \pmb{0.2132} & 7.24\%  \\
& & HR $\uparrow$     & 0.9623   & 0.9369 & 0.9481 & \underline{0.9767} & 0.9694   & 0.9694 & 0.9575 & \pmb{0.9863} & 0.98\%  \\
& & NDCG $\uparrow$   & 0.1528   & 0.1279 & 0.1396 & \underline{0.1914} & 0.1745   & 0.1731 & 0.1560 & \pmb{0.2102} & 9.82\%  \\ \hline

\end{tabular}}
\end{table*}

\subsection{Comparison of Experimental Results}
Table~\ref{Comparison} shows the experimental results of all methods on three real-world datasets. To more intuitively demonstrate the superiority of MCDCF, we calculated the improvement of each metric in MCDCF compared to the second-best one, using 'Imp.' to denote this degree of enhancement. Through comparison, the impact of MCDCF's improvement across various indicators is evident.

The comparison results indicate that our MCDCF method consistently achieves the best performance on each metric across the three real-world datasets, with the highest improvement of 80.00\% compared to the second-best result.

The effect of IPS-based debiasing methods is not stable. Specifically, the performance of IPS and its variant IPS-C is poor because they rely on item popularity to calculate the inverse propensity scores. When the popularity distribution among items varies significantly, the IPS debiasing method is prone to have higher variance, leading to unsatisfactory debiasing effects. Although the IPS-C method alleviates this problem by setting an upper limit on the inverse propensity score, this simple remedy has limited effect. Therefore, IPS-CN uses normalization, while IPS-CNSR further adds smoothing and renormalization to more effectively alleviate the high variance problem.
This enables IPS-CN and IPS-CNSR to achieve more stable debiasing effects.
However, in contrast to MCDCF, the IPS-CN and IPS-CNSR do not capture confounders. They simply balance the importance of popular and cold items without modelling confounders, and thus their performance improvement is limited.

While the DICE and DCCL methods can capture specific confounders, they rely on the assumption that the mechanisms underlying the data generation are consistent with their causal graph. If biases exist in the dataset that contradict these assumptions (i.e., if the specific confounders targeted by these methods are absent), the effectiveness of both methods is limited. In contrast, the MCDCF approach focuses on identifying and capturing latent confounders on both the user and item sides, offering greater flexibility in addressing a variety of complex data environments.

We do not include a comparison with DeConfounder and its variants~\cite{wang2020causal,zhang2023debiasing,zhu2024deep} because DeConfounder focuses on explicit recommendation tasks, while our approach targets implicit recommendations. Additionally, the variants of DeConfounder require supplementary user information, which is often difficult to obtain due to privacy concerns and is not available in our dataset.

\begin{table}[hbp]
\centering
\caption{Ablation study of MCDCF on the three datasets.}
\label{Ablation}
\setlength{\tabcolsep}{0.8mm}{
\begin{tabular}{cc|ccc}
\hline
Dataset & Method & Recall@50$\uparrow$           & HR@50$\uparrow$             & NDCG@50$\uparrow$             \\ \hline
\multirow{4}{*}{Douban-Movie} 
& MF      & 0.0351 & 0.1162 & 0.0172 \\
& MCDCF-U & 0.0429 & 0.1448 & 0.0216 \\
& MCDCF-I & 0.0478 & 0.1567 & 0.0228 \\
& MCDCF  & \pmb{0.0570} & \pmb{0.1820} & \pmb{0.0275} \\ \hline
\multirow{4}{*}{Amazon-Art}   
& MF      & 0.0180 & 0.0579 & 0.0073 \\
& MCDCF-U & 0.0337 & 0.1010 & 0.0149 \\
& MCDCF-I & 0.0324 & 0.0969 & 0.0142 \\
& MCDCF   & \pmb{0.0442} & \pmb{0.1286} & \pmb{0.0199} \\ \hline
\multirow{4}{*}{KGRec-Music}  
& MF      & 0.1597 & 0.9623 & 0.1528 \\
& MCDCF-U & 0.1983 & 0.9769 & 0.1916 \\
& MCDCF-I & 0.1866 & 0.9692 & 0.1776 \\
& MCDCF   & \pmb{0.2132} & \pmb{0.9863} & \pmb{0.2102} \\ \hline
\end{tabular}}
\end{table}

\subsection{Ablation Studies}
We performed ablation studies to assess the effectiveness of each component of the MCDCF. We propose two variants of MCDCF: MCDCF-U and MCDCF-I, which denote only $X_U$ and $X_I$ are recovered, respectively. 

Table~\ref{Ablation} shows the experimental results of MCDCF and its variants on the three real-world datasets. The experimental results show that both MCDCF-U and MCDCF-I improve recommendation effectiveness, confirming the effectiveness of latent confounders $X_U$ and $X_I$. Moreover, MCDCF outperforms both variants, demonstrating the effectiveness and superiority of its strategy for recovering latent confounders on both the user and item sides. The consistent results across three real-world datasets illustrate the robustness of the MCDCF method.

\subsection{Evaluation on Debiasing Ability}

We use the \textit{Intersection Over Union} (\textit{IOU})~\cite{zheng2021disentangling} metric to evaluate the debiasing effectiveness of each method, quantified by calculating the proportion of popular items in the recommended list. A higher IOU value indicates a greater proportion of popular items, corresponding to poorer debiasing effectiveness.

\begin{figure}[htbp]
  \centering
  \includegraphics[width=0.9\columnwidth]{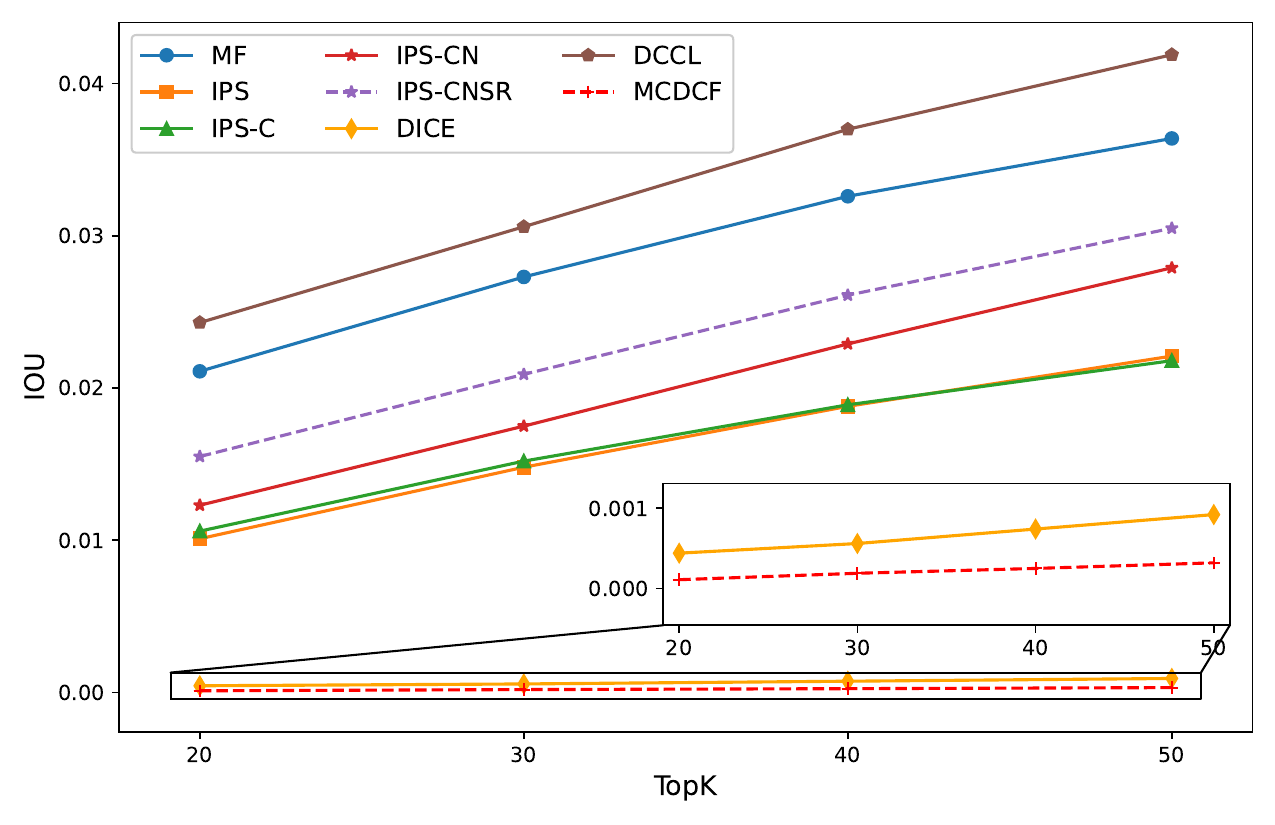}
  \caption{The IOU curves of all method on the Douban-Movie dataset.}
  \Description{}
  \label{Debiasing_Douban}
\end{figure}

In Figure~\ref{Debiasing_Douban}, we show the IOU curves of all methods on the Douban-Movie dataset. The results indicate that the IOU curve of MCDCF is the lowest, demonstrating that MCDCF exhibits the strongest debiasing ability among all compared methods. Meanwhile, the IOU curves of the other methods generally show an increasing trend when the number of recommended items increases, which reflects that their debiasing ability diminishes with the expansion of the recommendation list. In contrast, the debiasing ability of MCDCF remains stable even with large TOP-K values. Additionally, in the debiasing experiments on two other real datasets (the results are included in Appendix~\ref{more_debiasing}), the IOU curves of MCDCF also consistently remain the lowest. This consistent result further confirms that the debiasing ability of MCDCF is not accidental, but is an effective manifestation of its intrinsic mechanism. Therefore, these results demonstrate the superiority and effectiveness of our MCDCF method in terms of debiasing ability.

\subsection{Intervention Experiments}
To confirm that MCDCF can indeed effectively learn substitute latent confounders $X_U$ and $X_I$ from biased user behavioural data, we designed a series of intervention experiments. Specifically, we introduced different proportions of unbiased data in the training dataset as a way to observe the performance of MCDCF in terms of debiasing ability and recommendation performance.

\begin{figure}[htbp]
  \centering
  \includegraphics[width=0.7\columnwidth]{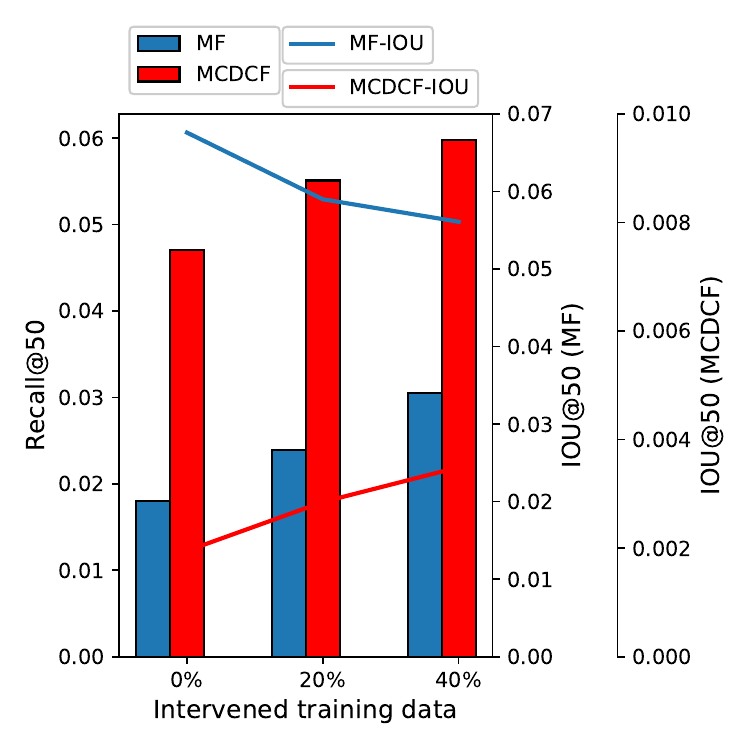}
  \caption{The IOU and Recall of MF and MCDCF on the Amazon-Art dataset.}
  \Description{}
  \label{Intervention}
\end{figure}

Figure~\ref{Intervention} illustrates the IOU and Recall of MF and MCDCF when different proportions of unbiased data are introduced in the Amazon-Art dataset. The recommendation performance of both MF and MCDCF is enhanced as the proportion of unbiased data increases. This enhancement occurs because the increase in unbiased data makes the training data distribution closer to the test data distribution, which improves the accuracy of the recommender system.

In addition, we notice that the IOU curve of MF shows a decreasing trend, which indicates that as the proportion of unbiased data in the training set increases, the model becomes less affected by bias, thereby enhancing its debiasing ability. In contrast, the IOU curve of MCDCF exhibits an increasing trend, which reflects that MCDCF is a process of learning to replace latent confounders based on biased data. As the amount of biased data decreases, the model's ability to capture information about the confounders becomes limited, resulting in a reduction in its debiasing effectiveness. The negative correlation between MCDCF's debiasing ability and the proportion of unbiased data further confirms its capability to extract latent confounders from biased data.

\section{Conclusion}
In this paper, we propose a novel multi-cause deconfounding method (MCDCF) for recommender systems in the presence of latent confounders. The MCDCF method separately learns substitute representations of latent confounders for both the user and item sides from user behaviour data. By conditioning click predictions on these factors, MCDCF aims to achieve a debiased recommender system. Unlike existing causal inference-based debiasing methods, MCDCF accounts for the influence of latent confounders on both the user and item sides, which represents a more challenging and general case. We conduct extensive experiments on three real-world datasets to validate the effectiveness and superiority of the MCDCF method in terms of both recommendation accuracy and debiasing performance.

\bibliographystyle{ACM-Reference-Format}
\bibliography{acmart}


\begin{thebibliography}{38}


\ifx \showCODEN    \undefined \def \showCODEN     #1{\unskip}     \fi
\ifx \showDOI      \undefined \def \showDOI       #1{#1}\fi
\ifx \showISBNx    \undefined \def \showISBNx     #1{\unskip}     \fi
\ifx \showISBNxiii \undefined \def \showISBNxiii  #1{\unskip}     \fi
\ifx \showISSN     \undefined \def \showISSN      #1{\unskip}     \fi
\ifx \showLCCN     \undefined \def \showLCCN      #1{\unskip}     \fi
\ifx \shownote     \undefined \def \shownote      #1{#1}          \fi
\ifx \showarticletitle \undefined \def \showarticletitle #1{#1}   \fi
\ifx \showURL      \undefined \def \showURL       {\relax}        \fi
\providecommand\bibfield[2]{#2}
\providecommand\bibinfo[2]{#2}
\providecommand\natexlab[1]{#1}
\providecommand\showeprint[2][]{arXiv:#2}

\bibitem[Bottou et~al\mbox{.}(2013)]%
        {bottou2013counterfactual}
\bibfield{author}{\bibinfo{person}{L{{\'e}}on Bottou}, \bibinfo{person}{Jonas
  Peters}, \bibinfo{person}{Joaquin Qui{{\~n}}onero-Candela},
  \bibinfo{person}{Denis~X. Charles}, \bibinfo{person}{D.~Max Chickering},
  \bibinfo{person}{Elon Portugaly}, \bibinfo{person}{Dipankar Ray},
  \bibinfo{person}{Patrice Simard}, {and} \bibinfo{person}{Ed Snelson}.}
  \bibinfo{year}{2013}\natexlab{}.
\newblock \showarticletitle{Counterfactual Reasoning and Learning Systems: The
  Example of Computational Advertising}.
\newblock \bibinfo{journal}{\emph{Journal of Machine Learning Research}}
  \bibinfo{volume}{14}, \bibinfo{number}{101} (\bibinfo{year}{2013}),
  \bibinfo{pages}{3207--3260}.
\newblock


\bibitem[Cai et~al\mbox{.}(2024)]%
        {cai2024mitigating}
\bibfield{author}{\bibinfo{person}{Miaomiao Cai}, \bibinfo{person}{Min Hou},
  \bibinfo{person}{Lei Chen}, \bibinfo{person}{Le Wu}, \bibinfo{person}{Haoyue
  Bai}, \bibinfo{person}{Yong Li}, {and} \bibinfo{person}{Meng Wang}.}
  \bibinfo{year}{2024}\natexlab{}.
\newblock \showarticletitle{Mitigating Recommendation Biases via
  Group-Alignment and Global-Uniformity in Representation Learning}.
\newblock \bibinfo{journal}{\emph{ACM Transactions on Intelligent Systems and
  Technology}} (\bibinfo{year}{2024}).
\newblock


\bibitem[Chen et~al\mbox{.}(2023)]%
        {chen2023bias}
\bibfield{author}{\bibinfo{person}{Jiawei Chen}, \bibinfo{person}{Hande Dong},
  \bibinfo{person}{Xiang Wang}, \bibinfo{person}{Fuli Feng},
  \bibinfo{person}{Meng Wang}, {and} \bibinfo{person}{Xiangnan He}.}
  \bibinfo{year}{2023}\natexlab{}.
\newblock \showarticletitle{Bias and debias in recommender system: A survey and
  future directions}.
\newblock \bibinfo{journal}{\emph{ACM Transactions on Information Systems}}
  \bibinfo{volume}{41}, \bibinfo{number}{3} (\bibinfo{year}{2023}),
  \bibinfo{pages}{1--39}.
\newblock


\bibitem[Cheng et~al\mbox{.}(2023)]%
        {cheng2023causal}
\bibfield{author}{\bibinfo{person}{Debo Cheng}, \bibinfo{person}{Ziqi Xu},
  \bibinfo{person}{Jiuyong Li}, \bibinfo{person}{Lin Liu},
  \bibinfo{person}{Jixue Liu}, {and} \bibinfo{person}{Thuc~Duy Le}.}
  \bibinfo{year}{2023}\natexlab{}.
\newblock \showarticletitle{Causal inference with conditional instruments using
  deep generative models}. In \bibinfo{booktitle}{\emph{Proceedings of the AAAI
  Conference on Artificial Intelligence}}, Vol.~\bibinfo{volume}{37}.
  \bibinfo{pages}{7122--7130}.
\newblock


\bibitem[Gao et~al\mbox{.}(2024)]%
        {gao2024causal}
\bibfield{author}{\bibinfo{person}{Chen Gao}, \bibinfo{person}{Yu Zheng},
  \bibinfo{person}{Wenjie Wang}, \bibinfo{person}{Fuli Feng},
  \bibinfo{person}{Xiangnan He}, {and} \bibinfo{person}{Yong Li}.}
  \bibinfo{year}{2024}\natexlab{}.
\newblock \showarticletitle{Causal inference in recommender systems: A survey
  and future directions}.
\newblock \bibinfo{journal}{\emph{ACM Transactions on Information Systems}}
  \bibinfo{volume}{42}, \bibinfo{number}{4} (\bibinfo{year}{2024}),
  \bibinfo{pages}{1--32}.
\newblock


\bibitem[Gomez-Uribe and Hunt(2015)]%
        {gomez2015netflix}
\bibfield{author}{\bibinfo{person}{Carlos~A Gomez-Uribe} {and}
  \bibinfo{person}{Neil Hunt}.} \bibinfo{year}{2015}\natexlab{}.
\newblock \showarticletitle{The netflix recommender system: Algorithms,
  business value, and innovation}.
\newblock \bibinfo{journal}{\emph{ACM Transactions on Management Information
  Systems (TMIS)}} \bibinfo{volume}{6}, \bibinfo{number}{4}
  (\bibinfo{year}{2015}), \bibinfo{pages}{1--19}.
\newblock


\bibitem[Gruson et~al\mbox{.}(2019)]%
        {gruson2019offline}
\bibfield{author}{\bibinfo{person}{Alois Gruson}, \bibinfo{person}{Praveen
  Chandar}, \bibinfo{person}{Christophe Charbuillet}, \bibinfo{person}{James
  McInerney}, \bibinfo{person}{Samantha Hansen}, \bibinfo{person}{Damien
  Tardieu}, {and} \bibinfo{person}{Ben Carterette}.}
  \bibinfo{year}{2019}\natexlab{}.
\newblock \showarticletitle{Offline evaluation to make decisions about
  playlistrecommendation algorithms}. In \bibinfo{booktitle}{\emph{Proceedings
  of the Twelfth ACM International Conference on Web Search and Data Mining}}.
  \bibinfo{pages}{420--428}.
\newblock


\bibitem[He et~al\mbox{.}(2023)]%
        {he2023addressing}
\bibfield{author}{\bibinfo{person}{Xiangnan He}, \bibinfo{person}{Yang Zhang},
  \bibinfo{person}{Fuli Feng}, \bibinfo{person}{Chonggang Song},
  \bibinfo{person}{Lingling Yi}, \bibinfo{person}{Guohui Ling}, {and}
  \bibinfo{person}{Yongdong Zhang}.} \bibinfo{year}{2023}\natexlab{}.
\newblock \showarticletitle{Addressing confounding feature issue for causal
  recommendation}.
\newblock \bibinfo{journal}{\emph{ACM Transactions on Information Systems}}
  \bibinfo{volume}{41}, \bibinfo{number}{3} (\bibinfo{year}{2023}),
  \bibinfo{pages}{1--23}.
\newblock


\bibitem[Hou et~al\mbox{.}(2024)]%
        {hou2024bridging}
\bibfield{author}{\bibinfo{person}{Yupeng Hou}, \bibinfo{person}{Jiacheng Li},
  \bibinfo{person}{Zhankui He}, \bibinfo{person}{An Yan},
  \bibinfo{person}{Xiusi Chen}, {and} \bibinfo{person}{Julian McAuley}.}
  \bibinfo{year}{2024}\natexlab{}.
\newblock \showarticletitle{Bridging language and items for retrieval and
  recommendation}.
\newblock \bibinfo{journal}{\emph{arXiv preprint arXiv:2403.03952}}
  (\bibinfo{year}{2024}).
\newblock


\bibitem[Kingma and Welling(2013)]%
        {kingma2013auto}
\bibfield{author}{\bibinfo{person}{Diederik~P Kingma} {and}
  \bibinfo{person}{Max Welling}.} \bibinfo{year}{2013}\natexlab{}.
\newblock \showarticletitle{Auto-encoding variational bayes}.
\newblock \bibinfo{journal}{\emph{arXiv preprint arXiv:1312.6114}}
  (\bibinfo{year}{2013}).
\newblock


\bibitem[Ko et~al\mbox{.}(2022)]%
        {ko2022survey}
\bibfield{author}{\bibinfo{person}{Hyeyoung Ko}, \bibinfo{person}{Suyeon Lee},
  \bibinfo{person}{Yoonseo Park}, {and} \bibinfo{person}{Anna Choi}.}
  \bibinfo{year}{2022}\natexlab{}.
\newblock \showarticletitle{A survey of recommendation systems: recommendation
  models, techniques, and application fields}.
\newblock \bibinfo{journal}{\emph{Electronics}} \bibinfo{volume}{11},
  \bibinfo{number}{1} (\bibinfo{year}{2022}), \bibinfo{pages}{141}.
\newblock


\bibitem[Koren et~al\mbox{.}(2009)]%
        {koren2009matrix}
\bibfield{author}{\bibinfo{person}{Yehuda Koren}, \bibinfo{person}{Robert
  Bell}, {and} \bibinfo{person}{Chris Volinsky}.}
  \bibinfo{year}{2009}\natexlab{}.
\newblock \showarticletitle{Matrix factorization techniques for recommender
  systems}.
\newblock \bibinfo{journal}{\emph{Computer}} \bibinfo{volume}{42},
  \bibinfo{number}{8} (\bibinfo{year}{2009}), \bibinfo{pages}{30--37}.
\newblock


\bibitem[Luo et~al\mbox{.}(2024)]%
        {luo2024survey}
\bibfield{author}{\bibinfo{person}{Huishi Luo}, \bibinfo{person}{Fuzhen
  Zhuang}, \bibinfo{person}{Ruobing Xie}, \bibinfo{person}{Hengshu Zhu},
  \bibinfo{person}{Deqing Wang}, \bibinfo{person}{Zhulin An}, {and}
  \bibinfo{person}{Yongjun Xu}.} \bibinfo{year}{2024}\natexlab{}.
\newblock \showarticletitle{A survey on causal inference for recommendation}.
\newblock \bibinfo{journal}{\emph{The Innovation}} (\bibinfo{year}{2024}).
\newblock


\bibitem[Oramas et~al\mbox{.}(2016)]%
        {oramas2016sound}
\bibfield{author}{\bibinfo{person}{Sergio Oramas},
  \bibinfo{person}{Vito~Claudio Ostuni}, \bibinfo{person}{Tommaso~Di Noia},
  \bibinfo{person}{Xavier Serra}, {and} \bibinfo{person}{Eugenio~Di Sciascio}.}
  \bibinfo{year}{2016}\natexlab{}.
\newblock \showarticletitle{Sound and music recommendation with knowledge
  graphs}.
\newblock \bibinfo{journal}{\emph{ACM Transactions on Intelligent Systems and
  Technology (TIST)}} \bibinfo{volume}{8}, \bibinfo{number}{2}
  (\bibinfo{year}{2016}), \bibinfo{pages}{1--21}.
\newblock


\bibitem[Papadakis et~al\mbox{.}(2022)]%
        {papadakis2022collaborative}
\bibfield{author}{\bibinfo{person}{Harris Papadakis}, \bibinfo{person}{Antonis
  Papagrigoriou}, \bibinfo{person}{Costas Panagiotakis},
  \bibinfo{person}{Eleftherios Kosmas}, {and} \bibinfo{person}{Paraskevi
  Fragopoulou}.} \bibinfo{year}{2022}\natexlab{}.
\newblock \showarticletitle{Collaborative filtering recommender systems
  taxonomy}.
\newblock \bibinfo{journal}{\emph{Knowledge and Information Systems}}
  \bibinfo{volume}{64}, \bibinfo{number}{1} (\bibinfo{year}{2022}),
  \bibinfo{pages}{35--74}.
\newblock


\bibitem[Pearl(2009)]%
        {pearl2009causality}
\bibfield{author}{\bibinfo{person}{Judea Pearl}.}
  \bibinfo{year}{2009}\natexlab{}.
\newblock \bibinfo{booktitle}{\emph{Causality}}.
\newblock \bibinfo{publisher}{Cambridge university press}.
\newblock


\bibitem[Rendle et~al\mbox{.}(2012)]%
        {rendle2012bpr}
\bibfield{author}{\bibinfo{person}{Steffen Rendle}, \bibinfo{person}{Christoph
  Freudenthaler}, \bibinfo{person}{Zeno Gantner}, {and} \bibinfo{person}{Lars
  Schmidt-Thieme}.} \bibinfo{year}{2012}\natexlab{}.
\newblock \showarticletitle{BPR: Bayesian personalized ranking from implicit
  feedback}.
\newblock \bibinfo{journal}{\emph{arXiv preprint arXiv:1205.2618}}
  (\bibinfo{year}{2012}).
\newblock


\bibitem[Schnabel et~al\mbox{.}(2016)]%
        {schnabel2016recommendations}
\bibfield{author}{\bibinfo{person}{Tobias Schnabel}, \bibinfo{person}{Adith
  Swaminathan}, \bibinfo{person}{Ashudeep Singh}, \bibinfo{person}{Navin
  Chandak}, {and} \bibinfo{person}{Thorsten Joachims}.}
  \bibinfo{year}{2016}\natexlab{}.
\newblock \showarticletitle{Recommendations as Treatments: Debiasing Learning
  and Evaluation}. In \bibinfo{booktitle}{\emph{Proceedings of The 33rd
  International Conference on Machine Learning}}, Vol.~\bibinfo{volume}{48}.
  \bibinfo{publisher}{PMLR}, \bibinfo{address}{New York, New York, USA},
  \bibinfo{pages}{1670--1679}.
\newblock


\bibitem[Sch{\"o}lkopf(2022)]%
        {scholkopf2022causality}
\bibfield{author}{\bibinfo{person}{Bernhard Sch{\"o}lkopf}.}
  \bibinfo{year}{2022}\natexlab{}.
\newblock \showarticletitle{Causality for machine learning}.
\newblock In \bibinfo{booktitle}{\emph{Probabilistic and causal inference: The
  works of Judea Pearl}}. \bibinfo{pages}{765--804}.
\newblock


\bibitem[Shoja and Tabrizi(2019)]%
        {shoja2019customer}
\bibfield{author}{\bibinfo{person}{Babak~Maleki Shoja} {and}
  \bibinfo{person}{Nasseh Tabrizi}.} \bibinfo{year}{2019}\natexlab{}.
\newblock \showarticletitle{Customer reviews analysis with deep neural networks
  for e-commerce recommender systems}.
\newblock \bibinfo{journal}{\emph{IEEE access}}  \bibinfo{volume}{7}
  (\bibinfo{year}{2019}), \bibinfo{pages}{119121--119130}.
\newblock


\bibitem[Si et~al\mbox{.}(2022)]%
        {si2022model}
\bibfield{author}{\bibinfo{person}{Zihua Si}, \bibinfo{person}{Xueran Han},
  \bibinfo{person}{Xiao Zhang}, \bibinfo{person}{Jun Xu}, \bibinfo{person}{Yue
  Yin}, \bibinfo{person}{Yang Song}, {and} \bibinfo{person}{Ji-Rong Wen}.}
  \bibinfo{year}{2022}\natexlab{}.
\newblock \showarticletitle{A model-agnostic causal learning framework for
  recommendation using search data}. In \bibinfo{booktitle}{\emph{Proceedings
  of the ACM Web Conference 2022}}. \bibinfo{pages}{224--233}.
\newblock


\bibitem[Si et~al\mbox{.}(2023a)]%
        {si2023enhancing}
\bibfield{author}{\bibinfo{person}{Zihua Si}, \bibinfo{person}{Zhongxiang Sun},
  \bibinfo{person}{Xiao Zhang}, \bibinfo{person}{Jun Xu}, \bibinfo{person}{Yang
  Song}, \bibinfo{person}{Xiaoxue Zang}, {and} \bibinfo{person}{Ji-Rong Wen}.}
  \bibinfo{year}{2023}\natexlab{a}.
\newblock \showarticletitle{Enhancing Recommendation with Search Data in a
  Causal Learning Manner}.
\newblock \bibinfo{journal}{\emph{ACM Transactions on Information Systems}}
  \bibinfo{volume}{41}, \bibinfo{number}{4} (\bibinfo{year}{2023}),
  \bibinfo{pages}{1--31}.
\newblock


\bibitem[Si et~al\mbox{.}(2023b)]%
        {si2023search}
\bibfield{author}{\bibinfo{person}{Zihua Si}, \bibinfo{person}{Zhongxiang Sun},
  \bibinfo{person}{Xiao Zhang}, \bibinfo{person}{Jun Xu},
  \bibinfo{person}{Xiaoxue Zang}, \bibinfo{person}{Yang Song},
  \bibinfo{person}{Kun Gai}, {and} \bibinfo{person}{Ji-Rong Wen}.}
  \bibinfo{year}{2023}\natexlab{b}.
\newblock \showarticletitle{When search meets recommendation: Learning
  disentangled search representation for recommendation}. In
  \bibinfo{booktitle}{\emph{Proceedings of the 46th International ACM SIGIR
  Conference on Research and Development in Information Retrieval}}.
  \bibinfo{pages}{1313--1323}.
\newblock


\bibitem[Singh et~al\mbox{.}(2021)]%
        {singh2021recommender}
\bibfield{author}{\bibinfo{person}{Pradeep~Kumar Singh},
  \bibinfo{person}{Pijush Kanti~Dutta Pramanik}, \bibinfo{person}{Avick~Kumar
  Dey}, {and} \bibinfo{person}{Prasenjit Choudhury}.}
  \bibinfo{year}{2021}\natexlab{}.
\newblock \showarticletitle{Recommender systems: an overview, research trends,
  and future directions}.
\newblock \bibinfo{journal}{\emph{International Journal of Business and Systems
  Research}} \bibinfo{volume}{15}, \bibinfo{number}{1} (\bibinfo{year}{2021}),
  \bibinfo{pages}{14--52}.
\newblock


\bibitem[Van~Erven and Harremos(2014)]%
        {van2014renyi}
\bibfield{author}{\bibinfo{person}{Tim Van~Erven} {and} \bibinfo{person}{Peter
  Harremos}.} \bibinfo{year}{2014}\natexlab{}.
\newblock \showarticletitle{R{\'e}nyi divergence and Kullback-Leibler
  divergence}.
\newblock \bibinfo{journal}{\emph{IEEE Transactions on Information Theory}}
  \bibinfo{volume}{60}, \bibinfo{number}{7} (\bibinfo{year}{2014}),
  \bibinfo{pages}{3797--3820}.
\newblock


\bibitem[Wang et~al\mbox{.}(2021)]%
        {wang2021non}
\bibfield{author}{\bibinfo{person}{Nan Wang}, \bibinfo{person}{Zhen Qin},
  \bibinfo{person}{Xuanhui Wang}, {and} \bibinfo{person}{Hongning Wang}.}
  \bibinfo{year}{2021}\natexlab{}.
\newblock \showarticletitle{Non-clicks mean irrelevant? propensity ratio
  scoring as a correction}. In \bibinfo{booktitle}{\emph{Proceedings of the
  14th ACM international conference on web search and data mining}}.
  \bibinfo{pages}{481--489}.
\newblock


\bibitem[Wang and Blei(2019)]%
        {wang2019blessings}
\bibfield{author}{\bibinfo{person}{Yixin Wang} {and} \bibinfo{person}{David~M
  Blei}.} \bibinfo{year}{2019}\natexlab{}.
\newblock \showarticletitle{The blessings of multiple causes}.
\newblock \bibinfo{journal}{\emph{J. Amer. Statist. Assoc.}}
  \bibinfo{volume}{114}, \bibinfo{number}{528} (\bibinfo{year}{2019}),
  \bibinfo{pages}{1574--1596}.
\newblock


\bibitem[Wang et~al\mbox{.}(2020)]%
        {wang2020causal}
\bibfield{author}{\bibinfo{person}{Yixin Wang}, \bibinfo{person}{Dawen Liang},
  \bibinfo{person}{Laurent Charlin}, {and} \bibinfo{person}{David~M Blei}.}
  \bibinfo{year}{2020}\natexlab{}.
\newblock \showarticletitle{Causal inference for recommender systems}. In
  \bibinfo{booktitle}{\emph{Proceedings of the 14th ACM Conference on
  Recommender Systems}}. \bibinfo{pages}{426--431}.
\newblock


\bibitem[Wei et~al\mbox{.}(2021)]%
        {wei2021model}
\bibfield{author}{\bibinfo{person}{Tianxin Wei}, \bibinfo{person}{Fuli Feng},
  \bibinfo{person}{Jiawei Chen}, \bibinfo{person}{Ziwei Wu},
  \bibinfo{person}{Jinfeng Yi}, {and} \bibinfo{person}{Xiangnan He}.}
  \bibinfo{year}{2021}\natexlab{}.
\newblock \showarticletitle{Model-agnostic counterfactual reasoning for
  eliminating popularity bias in recommender system}. In
  \bibinfo{booktitle}{\emph{Proceedings of the 27th ACM SIGKDD conference on
  knowledge discovery \& data mining}}. \bibinfo{pages}{1791--1800}.
\newblock


\bibitem[Wu et~al\mbox{.}(2022)]%
        {wu2022survey}
\bibfield{author}{\bibinfo{person}{Le Wu}, \bibinfo{person}{Xiangnan He},
  \bibinfo{person}{Xiang Wang}, \bibinfo{person}{Kun Zhang}, {and}
  \bibinfo{person}{Meng Wang}.} \bibinfo{year}{2022}\natexlab{}.
\newblock \showarticletitle{A survey on accuracy-oriented neural
  recommendation: From collaborative filtering to information-rich
  recommendation}.
\newblock \bibinfo{journal}{\emph{IEEE Transactions on Knowledge and Data
  Engineering}} \bibinfo{volume}{35}, \bibinfo{number}{5}
  (\bibinfo{year}{2022}), \bibinfo{pages}{4425--4445}.
\newblock


\bibitem[Zhang et~al\mbox{.}(2023)]%
        {zhang2023debiasing}
\bibfield{author}{\bibinfo{person}{Qing Zhang}, \bibinfo{person}{Xiaoying
  Zhang}, \bibinfo{person}{Yang Liu}, \bibinfo{person}{Hongning Wang},
  \bibinfo{person}{Min Gao}, \bibinfo{person}{Jiheng Zhang}, {and}
  \bibinfo{person}{Ruocheng Guo}.} \bibinfo{year}{2023}\natexlab{}.
\newblock \showarticletitle{Debiasing recommendation by learning identifiable
  latent confounders}. In \bibinfo{booktitle}{\emph{Proceedings of the 29th ACM
  SIGKDD Conference on Knowledge Discovery and Data Mining}}.
  \bibinfo{pages}{3353--3363}.
\newblock


\bibitem[Zhang et~al\mbox{.}(2021)]%
        {zhang2021causal}
\bibfield{author}{\bibinfo{person}{Yang Zhang}, \bibinfo{person}{Fuli Feng},
  \bibinfo{person}{Xiangnan He}, \bibinfo{person}{Tianxin Wei},
  \bibinfo{person}{Chonggang Song}, \bibinfo{person}{Guohui Ling}, {and}
  \bibinfo{person}{Yongdong Zhang}.} \bibinfo{year}{2021}\natexlab{}.
\newblock \showarticletitle{Causal intervention for leveraging popularity bias
  in recommendation}. In \bibinfo{booktitle}{\emph{Proceedings of the 44th
  international ACM SIGIR conference on research and development in information
  retrieval}}. \bibinfo{pages}{11--20}.
\newblock


\bibitem[Zhao et~al\mbox{.}(2022b)]%
        {zhao2022investigating}
\bibfield{author}{\bibinfo{person}{Minghao Zhao}, \bibinfo{person}{Le Wu},
  \bibinfo{person}{Yile Liang}, \bibinfo{person}{Lei Chen},
  \bibinfo{person}{Jian Zhang}, \bibinfo{person}{Qilin Deng},
  \bibinfo{person}{Kai Wang}, \bibinfo{person}{Xudong Shen},
  \bibinfo{person}{Tangjie Lv}, {and} \bibinfo{person}{Runze Wu}.}
  \bibinfo{year}{2022}\natexlab{b}.
\newblock \showarticletitle{Investigating accuracy-novelty performance for
  graph-based collaborative filtering}. In
  \bibinfo{booktitle}{\emph{Proceedings of the 45th international ACM SIGIR
  conference on research and development in information retrieval}}.
  \bibinfo{pages}{50--59}.
\newblock


\bibitem[Zhao et~al\mbox{.}(2023)]%
        {zhao2023disentangled}
\bibfield{author}{\bibinfo{person}{Weiqi Zhao}, \bibinfo{person}{Dian Tang},
  \bibinfo{person}{Xin Chen}, \bibinfo{person}{Dawei Lv},
  \bibinfo{person}{Daoli Ou}, \bibinfo{person}{Biao Li}, \bibinfo{person}{Peng
  Jiang}, {and} \bibinfo{person}{Kun Gai}.} \bibinfo{year}{2023}\natexlab{}.
\newblock \showarticletitle{Disentangled Causal Embedding With Contrastive
  Learning For Recommender System}. In \bibinfo{booktitle}{\emph{Companion
  Proceedings of the ACM Web Conference 2023}}. \bibinfo{publisher}{Association
  for Computing Machinery}, \bibinfo{address}{New York, NY, USA},
  \bibinfo{pages}{406–410}.
\newblock


\bibitem[Zhao et~al\mbox{.}(2022a)]%
        {zhao2022popularity}
\bibfield{author}{\bibinfo{person}{Zihao Zhao}, \bibinfo{person}{Jiawei Chen},
  \bibinfo{person}{Sheng Zhou}, \bibinfo{person}{Xiangnan He},
  \bibinfo{person}{Xuezhi Cao}, \bibinfo{person}{Fuzheng Zhang}, {and}
  \bibinfo{person}{Wei Wu}.} \bibinfo{year}{2022}\natexlab{a}.
\newblock \showarticletitle{Popularity bias is not always evil: Disentangling
  benign and harmful bias for recommendation}.
\newblock \bibinfo{journal}{\emph{IEEE Transactions on Knowledge and Data
  Engineering}} \bibinfo{volume}{35}, \bibinfo{number}{10}
  (\bibinfo{year}{2022}), \bibinfo{pages}{9920--9931}.
\newblock


\bibitem[Zheng et~al\mbox{.}(2021)]%
        {zheng2021disentangling}
\bibfield{author}{\bibinfo{person}{Yu Zheng}, \bibinfo{person}{Chen Gao},
  \bibinfo{person}{Xiang Li}, \bibinfo{person}{Xiangnan He},
  \bibinfo{person}{Yong Li}, {and} \bibinfo{person}{Depeng Jin}.}
  \bibinfo{year}{2021}\natexlab{}.
\newblock \showarticletitle{Disentangling user interest and conformity for
  recommendation with causal embedding}. In
  \bibinfo{booktitle}{\emph{Proceedings of the Web Conference 2021}}.
  \bibinfo{pages}{2980--2991}.
\newblock


\bibitem[Zhu et~al\mbox{.}(2024b)]%
        {zhu2024mitigating}
\bibfield{author}{\bibinfo{person}{Xinyuan Zhu}, \bibinfo{person}{Yang Zhang},
  \bibinfo{person}{Xun Yang}, \bibinfo{person}{Dingxian Wang}, {and}
  \bibinfo{person}{Fuli Feng}.} \bibinfo{year}{2024}\natexlab{b}.
\newblock \showarticletitle{Mitigating hidden confounding effects for causal
  recommendation}.
\newblock \bibinfo{journal}{\emph{IEEE Transactions on Knowledge and Data
  Engineering}} (\bibinfo{year}{2024}).
\newblock


\bibitem[Zhu et~al\mbox{.}(2024a)]%
        {zhu2024deep}
\bibfield{author}{\bibinfo{person}{Yaochen Zhu}, \bibinfo{person}{Jing Yi},
  \bibinfo{person}{Jiayi Xie}, {and} \bibinfo{person}{Zhenzhong Chen}.}
  \bibinfo{year}{2024}\natexlab{a}.
\newblock \showarticletitle{Deep causal reasoning for recommendations}.
\newblock \bibinfo{journal}{\emph{ACM Transactions on Intelligent Systems and
  Technology}} \bibinfo{volume}{15}, \bibinfo{number}{4}
  (\bibinfo{year}{2024}), \bibinfo{pages}{1--25}.
\newblock


\end{thebibliography}

\clearpage
\appendix

\section{Preliminaries}
\label{Preliminaries}
In this section, we will introduce the fundamental concepts of causal inference related to our main manuscript.

\subsection{Directed Acyclic Graph}
\label{Directed Acyclic Graph}
Causal graphs utilize Directed Acyclic Graphs (DAGs) to depict causal relationships among variables, where nodes represent variables and edges signify their causal connections. Three classical structures are commonly used to describe these relationships: the chain $A\rightarrow B \rightarrow C$, the fork $A\leftarrow B \rightarrow C$, and the collider $A\rightarrow B \leftarrow C$.

In the chain structure $A\rightarrow B \rightarrow C$, variable $A$ influences $C$ through the intermediary $B$. In the fork structure $A\leftarrow B \rightarrow C$, $B$ acts as a confounder, or common cause, of both $A$ and $C$. This means that $B$ influences both $A$ and $C$, creating a correlation between them; however, this correlation does not indicate a direct causal relationship. In the collider structure $A\rightarrow B \leftarrow C$, $A$ and $C$ are independent of each other but both influence the collider node $B$. When conditioned on $B$, a correlation arises between $A$ and $C$.


\subsection{D-Separation}
\label{D-Separation}
D-Separation is a fundamental concept in graphical models~\cite{pearl2009causality}, used to determine whether a set of variables is conditionally independent of another set given certain variables in a DAG. It serves as a crucial criterion for assessing conditional dependence or independence in such models.

\begin{definition}[$d$-separation~\cite{pearl2009causality}]
A path is said to be d-separated (or blocked) by conditioning on a set of nodes $Z$ if and only if one of the two conditions is satisfied:
\begin{enumerate}
    \item The $path$ contains a chain structure or a fork structure, and the middle node $B$ belongs to $Z$.
    \item The $path$ contains a collision structures and the collision node $B$ and its descendant nodes are not in $Z$.
\end{enumerate}
\end{definition}
\noindent where a $path$ represents a sequence of consecutive edges (directionless) in a DAG. 

\section{Preprocessing details of the dataset}
\label{Preprocessing}
We uniformly preprocessed each dataset to ensure the experiment's fairness and the results' comparability. First, we randomly selected 30\% of the data with the same probability (these data can be considered unbiased and represent the recommendation results under a completely random strategy). Then, we used 10\% of the unbiased data as the validation set and 20\% of the data as the test set. Overall, we set up the training, validation and test sets in the ratio of $7:1:2$, where the training set is biased data and the validation and test sets are unbiased data. 

\section{experimental parameter settings}
\label{Parameter}
To ensure fair comparisons, we standardised the parameter counts across all methods. The embedding size for models utilising DICE~\cite{zheng2021disentangling} and DCCL~\cite{zhao2023disentangled} was set to 64, as they consist of two concatenated sets of embeddings. For the other models, we maintained a consistent embedding size of 128. We set the batch size for DCCL to 4,096, while other methods use a batch size of 128. The larger batch size for DCCL is essential because contrastive learning approach requires a greater number of negative samples to effectively distinguish between positive and negative examples. In our  MCDCF method, we set the hyperparameters $\alpha$ and $\beta$ to $0.5$. We employed the Adam Optimiser to update model weights, with an initial learning rate of 0.01. For all baseline models except DICE and DCCL, the loss functions were based on the BPR~\cite{rendle2012bpr} function. All models were executed on an NVIDIA A100 (40GB RAM) GPU. To evaluate model performance, we used three widely recognised metrics in the field of recommender systems: \textit{Recall}, \textit{Hit Ratio} (\textit{HR}), and \textit{NDCG}.

\section{MORE EXPERIMENTAL RESULTS}
\label{more_debiasing}

\begin{figure}[hbp]
  \centering
  \includegraphics[width=0.8\columnwidth]{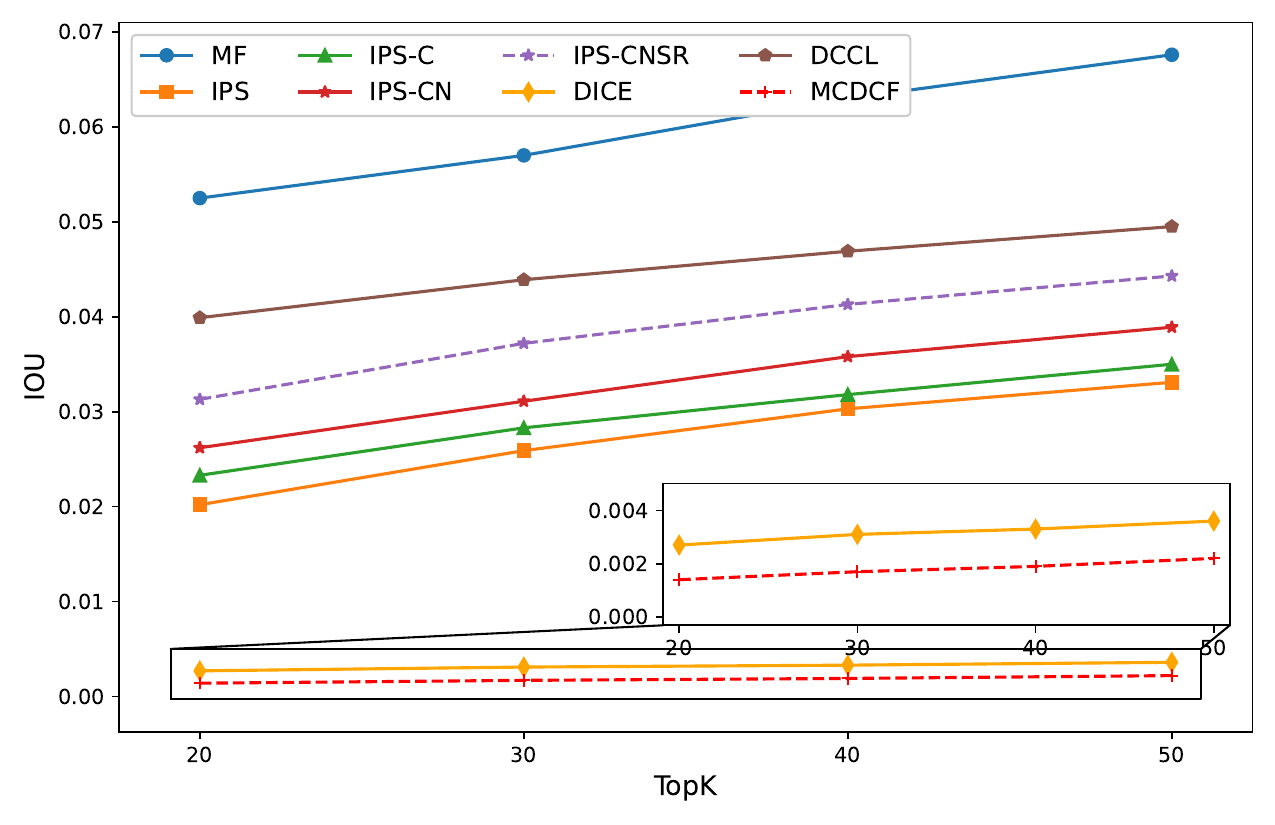}
  \caption{The IOU curves of each method on the Amazon-Art dataset.}
  \Description{}
  \label{Debiasing_art}
\end{figure}

\begin{figure}[hbp]
  \centering
  \includegraphics[width=0.8\columnwidth]{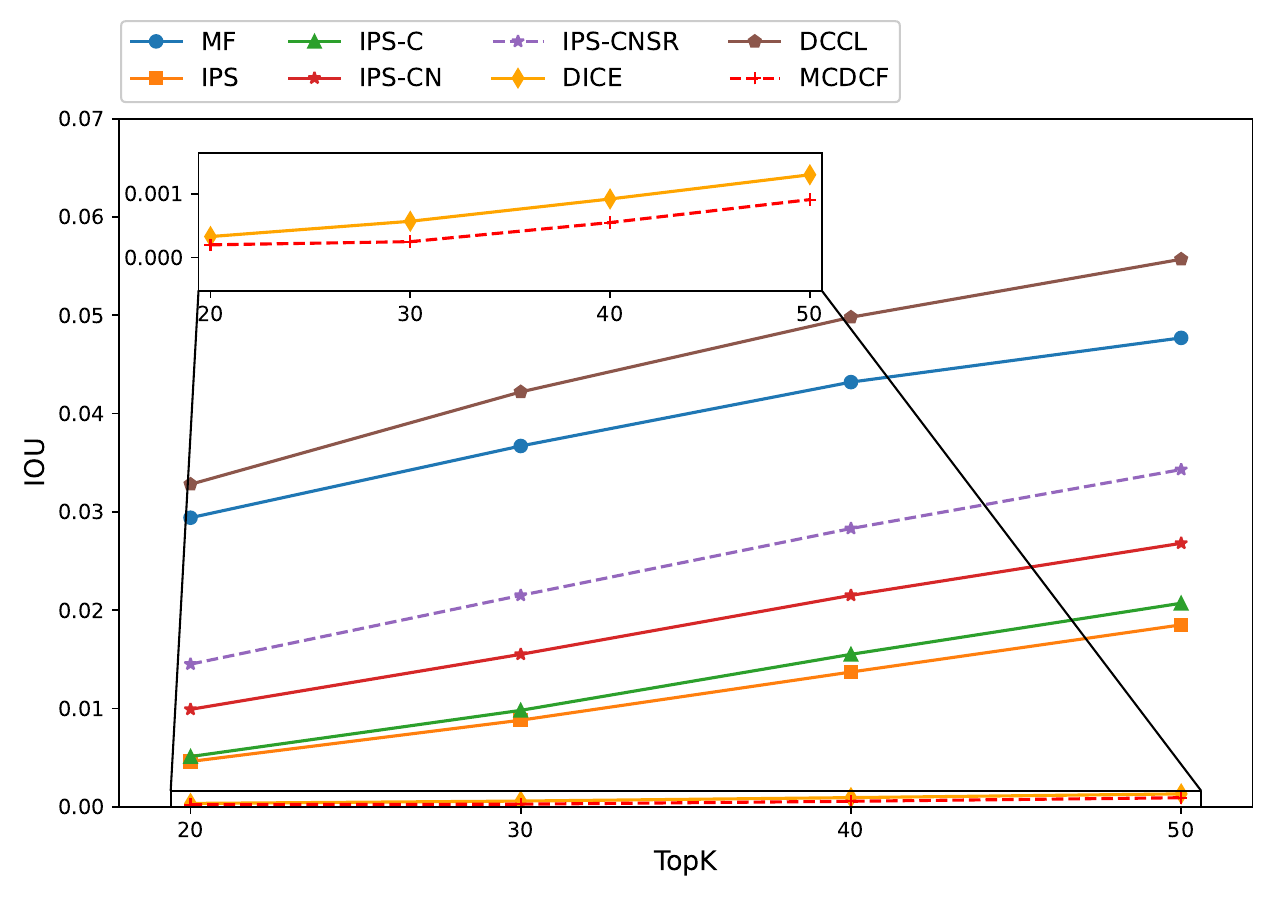}
  \caption{The IOU curves of each method on the KGRec-Music dataset.}
  \Description{}
  \label{Debiasing_Music}
\end{figure}
In Figures~\ref{Debiasing_art} and~\ref{Debiasing_Music}, we show the debiasing experimental results of the Amazon-Art and KGRec-Music datasets. MCDCF also achieved the best debiasing performance on the Amazon-Art and KGRec-Music datasets. MCDCF shows consistent debiasing performance on all three datasets. These findings suggest that MCDCF is not only effective in mitigating bias but also exhibits excellent adaptability to different dataset characteristics, such as different degrees of sparsity and latent confounders. The consistent results across different datasets confirm the superiority and effectiveness of MCDCF.
\end{document}